\documentstyle[prc,preprint,aps]{revtex}

\tightenlines

\begin{document}

\draft

\title{
Model-independent determination of the
$^{12}$C{\boldmath $(p,p')$}$^{12}$C$^{\ast}$ (15.11~MeV, 1$^+$, 
{\boldmath $T$}=1) transition amplitude at 200~MeV }

\author{
S. P. Wells\cite{spw} and S. W. Wissink }

\address{
Indiana University Cyclotron Facility, Bloomington, Indiana 47408 }

\date{\today}
\maketitle

\begin{abstract}
Using data obtained through simultaneous measurements of $(\vec p,\vec
p\,')$ spin-transfer observables and $(\vec p,p' \gamma)$ coincident
spin observables, we have made a model-independent determination of
the complete scattering amplitude for the 15.11~MeV, 1$^+$, $T$=1
state in $^{12}$C at an incident proton energy of 200~MeV, for four
proton scattering angles ranging from $\theta_{\rm c.m.}$ =
5.5$^\circ$ to 16.5$^\circ$. At each angle, 16 different observables
were determined, whereas only 11 independent quantities are required
to specify the transition amplitude for this state. It had been shown
previously that the set of observables measured span the allowed
space; hence the system is overdetermined, which allowed us to
extract, in a model-independent fashion, each of the individual
spin-operator amplitudes that characterize the reaction. Additional
insight into the physical mechanisms that drive this transition is
obtained by mapping out the momentum-transfer dependence of these
amplitudes. We also compare the magnitudes and phases determined for
each of the spin-operator amplitudes to the predictions of
calculations performed in both relativistic and nonrelativistic
frameworks, and discuss the physics content of these comparisons.
\end{abstract}



\section{Introduction}

At intermediate energies of $\sim$150--500~MeV, hadron-induced nuclear
reactions have served as a rich source of information, both to further
our understanding of nuclear structure, and to illustrate how the
elementary nucleon-nucleon ($NN$) interaction may be modified in the
nuclear medium \cite{Bag92}. This is particularly true of spin
observables, which often contain contributions from interferences
between different pieces of the full scattering amplitude. These
interferences, inaccessible through measurements of differential cross
sections alone, can thus provide information about not only the
magnitudes, but also the relative phases of individual terms in the
transition amplitude. Ideally, one would like to extract precise
values for each of these (complex) terms in a manner which does not
rely on specific model assumptions. This information, in turn, can
then provide the most stringent tests of a given theoretical
prediction for the hadronic process of interest.

To obtain deeper insight into the nucleon-nucleus ($NA$) interaction,
several inelastic transitions have been identified as being
particularly amenable to experimental investigation. The strong,
isovector 1$^+$ state in $^{12}$C at an excitation energy of 15.11~MeV
has been extensively studied, using a variety of probes operating
under a broad range of kinematic conditions. Due to its $1^+$ nature,
hadronic excitation of this state represents an ``unnatural parity''
transition; consequently, this process should be particularly
sensitive to the spin-dependent parts of the $NA$ interaction, and in
fact has long served as a critical test for our understanding of the
$\Delta S = \Delta T = 1$ component of the $NN$ effective interaction
\cite{Lov85}. As an example, the spin observable $P-A_y$, i.e., the
difference between the polarization induced in the outgoing proton and
the scattering yield asymmetry that results from use of a polarized
incident beam, has been investigated both experimentally
\cite{Car82,Hic88} and theoretically \cite{Lov84}. This quantity,
which in a non-relativistic framework contains only interference terms
between competing pieces of the transition amplitude, has been shown
to be sensitive to the coupling of the nucleon spin to the bound
nucleon current \cite{Lov84}. In later relativistic treatments of
proton-nucleus scattering \cite{Pik85,Shep86}, it became clear that
these same nuclear current terms appear more naturally in a
relativistic formalism, and are produced through the linear couplings
between the upper and lower components of the bound nucleon wave
function \cite{Pik85,Shep86}. In either description, $P-A_y$ is
dependent on the momentum of the nucleons inside the nucleus, and is
therefore sensitive to the off-shell behavior of these nucleons, and
the non-local or exchange nature of the interaction.

Thus, detailed comparisons of measured spin observable data to various
predictions for specific nuclear transitions has traditionally been
our best means of constraining theoretical models.  In recent years,
though, such comparisons have raised a number of concerns, at both the
experimental and theoretical ends. For example, it has been shown
\cite{Mos82,Ble82} that in a direct-only plane-wave impulse
approximation (PWIA), certain combinations of $(\vec p,\vec p\,')$
spin-transfer observables are directly related to individual terms in
the effective $NN$ interaction, weighted by corresponding nuclear
response functions. When physical processes such as distortion of the
projectile waves and knock-on exchange are included, however, it
becomes less clear what these combinations of $(\vec p,\vec p\,')$
observables represent physically, thereby blurring any simple
interpretation of these quantities. On the experimental side, counting
arguments show that even ``complete'' sets of $(\vec p,\vec p\,')$
spin-transfer measurements provide (at most) eight independent pieces
of information, while a transition of the general form $0^+ \to
J^{\pi}$ requires knowledge of $(8J+3)$ independent quantities in
order to fully specify the scattering amplitude.

To reduce this ambiguity, it is clearly desirable to establish a more
direct connection between theory and experiment, such as would be
obtained from a model-independent extraction of individual elements of
the scattering amplitude. In so doing, one bypasses the `intermediate'
role played by spin observables, which represent non-trivial
combinations of many of these elements.  Carrying out such a program,
though, requires access to information beyond that provided by $(\vec
p,\vec p\,')$ spin-transfer measurements alone. In particular, one
must probe the polarization state of the `other particle' involved in
the reaction: the recoil nucleus. This can be achieved through study
of the angular correlation in the final state between the scattered
(out-going) proton and the particle(s) emitted in the decay of the
excited nucleus, as in reactions of the type $(\vec p,p' \gamma)$.
More specifically, for proton excitation of a $0^+ \to 1^+$
transition, followed by electromagnetic decay back to the ground
state, it has recently been proven formally \cite{Pik90,Ram94} that
certain $(\vec p,p' \gamma )$ measurements, when combined with
complete sets of $(\vec p,\vec p\,')$ observables, allow for a
complete description of the transition amplitude. Note that in this
case a total of $8J+3=11$ independent quantities must be determined.

In this paper, we report the first model-independent determination of
the complete transition amplitude for any nuclear final state with $J
\neq 0$. This analysis is based on a detailed set of measurements
\cite{SPW95} carried out at the Indiana University Cyclotron Facility
(IUCF), in which both $(\vec p,\vec p\,')$ spin-transfer and $(\vec
p,p' \gamma)$ coincident observables were determined simultaneously
for excitation of the isovector $1^+$ state at 15.11~MeV in $^{12}$C.
Data were taken at four proton scattering angles ($\theta_{\rm c.m.}$
= 5.5$^\circ$, 8.8$^\circ$, 12.1$^\circ$, and 16.5$^\circ$) at an
incident proton beam energy of 200~MeV. As will be shown below, this
data set allows for extraction of each individual (complex)
spin-dependent term in the transition amplitude in a model-independent
manner.  Moreover, because this procedure has been applied at a number
of scattering angles, the momentum-transfer ($q$) dependence of each
spin-operator amplitude has been mapped out, allowing for a clearer
interpretation of the physical mechanisms that drive this transition.

This paper will be organized in the following way. In Sec.~II we
present the formalism adopted (from Ref.~\cite{Pik90}) for analysis of
the data. In Sec.~III, we present the observables measured in the IUCF
experiment \cite{SPW95}, and express each of these in terms of real
and imaginary parts of the individual spin-operator amplitudes. We
describe in detail the procedure used to extract these amplitudes from
the observables in Sec.~IV, and present the results of this procedure
in Sec.~V. Section~VI provides a detailed discussion of the
momentum-transfer dependence determined for each amplitude (both
magnitude and phase), and compares our results to predictions of
calculations performed in both relativistic and nonrelativistic
formalisms. Where possible, we discuss the specific physics issues to
which each amplitude is sensitive. Our most significant results and
conclusions are summarized in Sec.~VII.

\section{Formalism}

The formalism adopted here follows that of Piekarewicz {\it et al.}
\cite{Pik90}, and has been presented in detail in Ref.~\cite{SPW95};
only the most important results are given here for completeness. We
work in the orthogonal coordinate system defined by
\begin{equation}
   {\bf n} \equiv {\bf p} \times {\bf p'}~; \qquad {\bf K} \equiv 
   {\bf p} + {\bf p'}~; \qquad {\bf q} \equiv {\bf n} \times {\bf K}~,
\label{eq:coord}
\end{equation}
where {\bf p} ({\bf p}$'$) is the incident (outgoing) proton momentum,
{\bf n} is directed normal to the scattering plane, {\bf K} is along
the direction of the average proton momentum, and (neglecting the
reaction $Q$-value) {\bf q} points in the direction of momentum
transfer ${\bf p^{\prime}}$--${\bf p}$. In this frame, the most
general form of the scattering amplitude allowed by angular momentum
and parity conservation for a $0^+ \to 1^+$ transition can be written
in the form \cite{Pik90}
\begin{eqnarray}
   \hat T ^p ({\bf p},{\bf p'}) 
  &~=~& A_{n0} ({\bf \hat \Sigma} \cdot {\bf \hat n}) 
   ~+~  A_{nn} ({\bf \hat \Sigma} \cdot {\bf \hat n})
                    ({\bf \sigma} \cdot {\bf \hat n}) 
   ~+~  A_{KK} ({\bf \hat \Sigma} \cdot {\bf \hat K})
                    ({\bf \sigma} \cdot {\bf \hat K}) \nonumber \\
  &~+~& A_{Kq} ({\bf \hat \Sigma} \cdot {\bf \hat K})
                    ({\bf \sigma} \cdot {\bf \hat q}) 
   ~+~  A_{qK} ({\bf \hat \Sigma} \cdot {\bf \hat q})
                    ({\bf \sigma} \cdot {\bf \hat K}) 
   ~+~  A_{qq} ({\bf \hat \Sigma} \cdot {\bf \hat q})
                    ({\bf \sigma} \cdot {\bf \hat q}) ~,
\label{eq:Tpp}
\end{eqnarray}
where 
\begin{equation}
   {\bf \hat \Sigma} _M ~\equiv~ 
   \vert 1^+ M \rangle \langle 0^+ \vert 
\end{equation}
is the polarization operator for promoting a $0^+$ state to a state
with $J^{\pi}$=$1^+$ and magnetic substate $M$, while ${\vec \sigma}$
are the Pauli spin operators for the projectile. In Eq.~\ref{eq:Tpp},
the $A_{i\mu}$'s are scalar functions of energy and momentum transfer,
where the subscripts $i (=n,K,q)$ and $\mu (=0,n,K,q)$ indicate the
polarization components of the recoil $1^+$ nucleus ($\bf \Sigma$) and
the scattered proton ($\vec \sigma$, with $\sigma _0 \equiv 1$),
respectively. Because there are six allowed complex amplitudes, 11
pieces of information (after eliminating one overall phase) are
required to specify the scattering amplitude for this transition.

If the final polarization of the nuclear state is undetected, one can
sum incoherently over the $\bf \Sigma$ index. In this case, the
relevant spin observables can be expressed as
\begin{eqnarray}
   {{d\sigma}\over{d\Omega _p}} 
   &~=~& \sum _{ij\mu} A_{i\mu}A^{\ast}_{j\mu} \delta _{ij} ~,
   \nonumber \\
   {{d\sigma}\over{d\Omega _p}}D_{\alpha \beta} 
   &~=~& \sum _{ij\mu \nu} A_{i\mu}A^{\ast}_{j\nu} \delta _{ij}
   {1\over 2}Tr \{ \sigma _{\alpha} \sigma _\mu \sigma _{\beta} 
   \sigma _{\nu} \} ~,
\label{eq:Dab}
\end{eqnarray}
where $\alpha ,\beta = 0,n,K,q.$ Using the scattering amplitude of
Eq.~\ref{eq:Tpp} and carrying through the Pauli algebra, it is easily
shown that only eight of the 16 possible singles $(\vec p,\vec p\,')$
spin observables (the $D_{\alpha \beta}$) are nonzero. Because 11
independent quantities appear in $\hat T ^p ({\bf p},{\bf p'})$, one
sees that singles measurements alone can not uniquely define this
amplitude; or, put another way, there is information contained in
$\hat T ^p ({\bf p},{\bf p'})$ which is not accessible via $(\vec
p,\vec p\,')$ observables. In particular, the presence in
Eq.~\ref{eq:Dab} of the Kronecker $\delta _{ij}$ makes it impossible
to determine the relative phase between any two spin operator
amplitudes ($A_{i\mu}$'s) that correspond to orthogonal orientations
of the recoil nuclear polarization.

If one makes the assumption that the $(p,p' \gamma)$ reaction is
strictly a two-step process, then the total transition amplitude
(excitation plus decay) can be written as the product of the strong
and electromagnetic amplitudes. In this case, all of the coincident
$(p,p' \gamma)$ observables may be written in terms of just the
singles $(p,p')$ spin operator amplitudes, $A_{i\mu}$, and the
$\gamma$-ray branching ratio to the ground state $R$ \cite{Pik90}. In
complete analogy with the singles observables (Eq.~\ref{eq:Dab}), the
spin-dependent coincident observables for this transition can be
written in the form \cite{Pik90}
\begin{eqnarray}
   {{d^2 \sigma}\over{d\Omega_p d\Omega_{\gamma}}}({\bf \hat k}) 
   &~=~& {{3R}\over{8 \pi}} \sum _{ij \mu} 
   A_{i \mu} A^{\ast}_{j \mu} t_{ij}({\bf \hat k}) ~, \nonumber \\
   {{d^2 \sigma}\over{d\Omega_p d\Omega_{\gamma}}}({\bf \hat k})
   D_{\alpha \beta} ({\bf \hat k}) 
   &~=~& \sum _{ij\mu\nu} A_{i\mu} A^{\ast}_{j\nu} t_{ij} ({\bf \hat k})
   \, {1\over 2}Tr \{ \sigma_{\alpha} \sigma_\mu \sigma_{\beta} 
   \sigma_{\nu} \} ~,
\label{eq:Dabk}
\end{eqnarray}
where ${\bf \hat k}$ is the momentum direction of the emitted photon,
and the photon polarization tensor $t_{ij}$ is given by
\begin{equation}
   t_{ij} ({\bf \hat k} ) ~\equiv~ 
   \delta_{ij} - ({\bf \hat k} \cdot {\bf \hat e}_i)
   ({\bf \hat k} \cdot {\bf \hat e}_j) ~,
\label{eq:tij}
\end{equation}
with ${\bf \hat e}_i$ a unit vector lying along one of the $({\bf \hat
n},{\bf \hat K},{\bf \hat q})$ coordinates defined in
Eq.~\ref{eq:coord}. It is important to emphasize that the coincident
observables defined in Eq.~\ref{eq:Dabk} are written in terms of the
{\em same\/} $A_{i\mu}$'s that appear in the definitions of the
singles observables, Eq.~\ref{eq:Dab}. The crucial difference between
these two sets of equations is the presence of $t_{ij} ({\bf \hat k}
)$ in the definition of the coincident observables, replacing the
$\delta _{ij}$ for the singles observables. Thus, certain $(p,p'
\gamma)$ spin observables {\em will\/} be sensitive to the relative
phases between amplitudes for different recoil nuclear polarizations,
provided that the emitted $\gamma$--ray has momentum components along
{\em both\/} ${\bf \hat e}_i$ and ${\bf \hat e}_j$.  Thus, the
specific information about the transition amplitude which is lost for
the singles observables can be accessed through coincident $(\vec p,p'
\gamma)$ measurements.

\section{Summary of Observables}

In this paper, we present detailed analysis of the data from a
previous experiment \cite{SPW95}. In that work, some spin observables
were measured using an incident proton beam whose polarization vector
was normal to the (horizontal) scattering plane, while others required
use of a beam in which the polarization vector had been precessed to
lie in the scattering plane. Due to technical problems encountered
during data acquisition \cite{SPW95}, reliable values of the
differential cross section, $d\sigma / d\Omega_p$, could not be
extracted from the experimental yields. While this problem had no
effect on the determination of spin-dependent observables, it was
necessary to use previously measured values of $d\sigma / d\Omega_p$
for this transition \cite{Com81} to set the scale for the overall
magnitudes of the $A_{i\mu}$ amplitudes. In this paper, these
amplitudes will be presented in units of $\sqrt{\mu {\rm b} / {\rm
sr}}$.

The singles observables obtained in Ref.~\cite{SPW95} include the
three normal-component spin-transfer coefficients $A_y$, $P$, and
$D_{N'N}$, and two linear combinations of the in-plane spin transfer
coefficients,
\begin{eqnarray}
   D_{\lambda} & \equiv & 
   D_{L'L} \sin \alpha + D_{S'L} \cos \alpha ~, \nonumber \\
   D_{\sigma} & \equiv & 
   D_{L'S} \sin \alpha + D_{S'S} \cos \alpha ~,
\end{eqnarray}
where here $\alpha$ is the spin precession angle about ${\bf \hat n}$
experienced by the scattered protons in passing through the magnetic
spectrometer used in the experiment. Through use of Eq.~\ref{eq:Dab},
all of these measured $(\vec p,\vec p\,')$ observables can be
expressed in terms of the $A_{i\mu}$ amplitudes as follows:
\begin{eqnarray}
   {{d\sigma}\over{d\Omega _p}} &~=~& 
   \vert A_{n0} \vert^2 + \vert A_{nn} \vert^2 + \vert A_{KK} \vert^2 
 + \vert A_{Kq} \vert^2 + \vert A_{qK} \vert^2 + \vert A_{qq} \vert^2
   ~, \nonumber \\
   {{d\sigma}\over{d\Omega _p}} A_y &~=~& 
   2[\Re (A_{n0}A^{\ast}_{nn}) +
   \Im (A_{KK}A^{\ast}_{Kq} + A_{qK}A^{\ast}_{qq})] ~, \nonumber \\
   {{d\sigma}\over{d\Omega _p}} P &~=~& 
   2[\Re (A_{n0}A^{\ast}_{nn}) -
   \Im (A_{KK}A^{\ast}_{Kq} + A_{qK}A^{\ast}_{qq})] ~, \nonumber \\
   {{d\sigma}\over{d\Omega _p}} D_{N'N} &~=~& 
   \vert A_{n0} \vert^2 + \vert A_{nn} \vert^2 - \vert A_{KK} \vert^2 
 - \vert A_{Kq} \vert^2 - \vert A_{qK} \vert^2 - \vert A_{qq} \vert^2 ~; 
\label{eq:norm}
\end{eqnarray}
and
\begin{eqnarray}
   {{d\sigma}\over{d\Omega _p}} D_{\lambda} &~=~& \cos \theta_{pK}
   [\sin (\alpha - \theta_{\rm c.m.} + \theta_{pK})
   {{d\sigma}\over{d\Omega _p}} D_{KK} + 
   \cos (\alpha - \theta_{\rm c.m.} + \theta_{pK})
   {{d\sigma}\over{d\Omega _p}}D_{qK}] \nonumber \\
   &~-~ & \sin \theta_{pK}
   [\sin (\alpha - \theta_{\rm c.m.} + \theta_{pK})
   {{d\sigma}\over{d\Omega _p}} D_{Kq} + 
   \cos (\alpha - \theta_{\rm c.m.} + \theta_{pK})
   {{d\sigma}\over{d\Omega _p}} D_{qq}] ~, \nonumber \\
   {{d\sigma}\over{d\Omega _p}} D_{\sigma} &~=~& \sin \theta _{pK}
   [\sin (\alpha - \theta_{\rm c.m.} + \theta_{pK})
   {{d\sigma}\over{d\Omega_p}} D_{KK} + 
   \cos (\alpha - \theta_{\rm c.m.} + \theta_{pK})
   {{d\sigma}\over{d\Omega _p}} D_{qK}] \nonumber \\
   &~+~& \cos \theta_{pK}
   [\sin (\alpha - \theta_{\rm c.m.} + \theta_{pK})
   {{d\sigma}\over{d\Omega _p}} D_{Kq} + 
   \cos (\alpha - \theta_{\rm c.m.} + \theta_{pK})
   {{d\sigma}\over{d\Omega _p}} D_{qq}] ~, 
\label{eq:Dsig}
\end{eqnarray}
where $\theta_{\rm c.m.}$ is the proton center-of-mass scattering
angle, $\theta_{pK}$ is the angle between the incident beam momentum
${\bf p}$ and the average momentum ${\bf K}$ (see Eq.~\ref{eq:coord}),
and $\alpha$ is the spin precession angle defined above. The
$D_{\alpha \beta}$ spin transfer coefficients that appear in the
previous equation can also be expressed in terms of the spin operator
amplitudes as:
\begin{eqnarray}
   {{d\sigma}\over{d\Omega_p}} D_{KK} &~=~& 
   \vert A_{n0} \vert^2 - \vert A_{nn} \vert^2 + \vert A_{KK} \vert^2 
 - \vert A_{Kq} \vert^2 + \vert A_{qK} \vert^2 - \vert A_{qq} \vert^2 
   ~, \nonumber \\
   {{d\sigma}\over{d\Omega_p}} D_{Kq} &~=~&
   -2[\Im (A_{n0}A^{\ast}_{nn}) -
   \Re (A_{KK}A^{\ast}_{Kq} + A_{qK}A^{\ast}_{qq})] ~, \nonumber \\
   {{d\sigma}\over{d\Omega_p}} D_{qK} &~=~& 
   2[\Im (A_{n0}A^{\ast}_{nn}) +
   \Re (A_{KK}A^{\ast}_{Kq} + A_{qK}A^{\ast}_{qq})] ~, \nonumber \\
   {{d\sigma}\over{d\Omega_p}} D_{qq} &~=~& 
   \vert A_{n0} \vert^2 - \vert A_{nn} \vert^2 - \vert A_{KK} \vert^2 
 + \vert A_{Kq} \vert^2 - \vert A_{qK} \vert^2 + \vert A_{qq} \vert^2 ~.
\label{eq:Dij}
\end{eqnarray}

For the coincident $(\vec p,p' \gamma)$ measurements, the incident
beam polarization was either pointed along the normal ${\bf \hat n}$
to the scattering plane, or was rotated to lie {\em in\/} the
scattering plane. During vertical polarization running, three photon
detectors were positioned in the scattering plane on beam right, and
one was placed directly above the target \cite{SPW95}. If a sum over
the two beam spin states was performed (effectively producing the
results that would be obtained with an unpolarized beam), the
coincident yields from the three in-plane detectors could be used to
extract the in-plane double-differential cross section, which takes
the form \cite{Pik90}
\begin{equation}
   {{8\pi}\over{3R}} {{d^2 \sigma}\over{d\Omega_{\gamma}d\Omega_p}} 
   (\theta_{\gamma}) ~=~ A(\theta_p) + B(\theta_p) \cos 2\theta_{\gamma} 
   + C(\theta_p) \sin 2\theta_{\gamma} ~,
\label{eq:d2sig}
\end{equation}
where $R$ is again the $\gamma$-ray branching ratio to the ground
state, $\theta_{\gamma}$ is the photon angle in the scattering plane,
measured with respect to the ${\bf \hat q}$ direction, and $A$, $B$,
and $C$ are unknown functions of the proton scattering angle. These
can be written in terms of amplitudes as:
\begin{eqnarray}
   A &~=~& \vert A_{n0} \vert^2 + \vert A_{nn} \vert^2 + {1 \over 2} 
   [\vert A_{KK} \vert^2 + \vert A_{Kq} \vert^2 + \vert A_{qK} \vert^2
   + \vert A_{qq} \vert^2] ~, \nonumber \\
   B &~=~& {1 \over 2} [\vert A_{KK} \vert^2 + \vert A_{Kq} \vert^2 
   - \vert A_{qK} \vert^2 - \vert A_{qq} \vert^2] ~, \nonumber \\
   C &~=~& -\Re [A_{KK}A^{\ast}_{qK} + A_{Kq}A^{\ast}_{qq}] ~.
\label{eq:ABC}
\end{eqnarray}
We point out that in this work our definitions of $A$, $B$, and $C$ do
not include the branching ratio normalization factor $8\pi / (3R)$,
and are therefore different from the definitions given in
Ref.~\cite{Pik90}. A more important comment is that, in order to
eliminate any sensitivity to detection efficiency in the photon
detectors, only the {\em ratios\/} of the symmetric and antisymmetric
pieces of the in-plane coincident cross section, $B/A$ and $C/A$, were
used in our determination of the transition amplitude. These
quantities have the obvious advantage of being independent of any
overall normalization error.

We now turn to the spin-dependent coincident observables which, at a
given proton scattering angle $\theta_p$, will be functions of the
angle $\theta_\gamma$ at which the photon is emitted. If the photon is
emitted in the scattering plane, then the normal-component coincident
analyzing power, scaled by the coincident cross section, can be cast
in a form similar to that of Eq.~\ref{eq:d2sig}, i.e.,
\begin{equation}
   {{8\pi}\over{3R}} {{d^2 \sigma} \over {d\Omega_{\gamma} d\Omega_p}}
   (\theta_{\gamma}) A_y (\theta_{\gamma}) ~=~ \epsilon_A (\theta_p ) + 
   \epsilon_B (\theta_p ) \cos 2\theta_{\gamma} +
   \epsilon_C (\theta_p ) \sin 2\theta_{\gamma} ~,
\end{equation}
where
\begin{eqnarray}
   \epsilon_A &~=~& 2\Re (A_{n0}A^{\ast}_{nn}) + 
   \Im (A_{KK}A^{\ast}_{Kq} + A_{qK}A^{\ast}_{qq}) ~, \nonumber \\
   \epsilon_B &~=~& \Im (A_{KK}A^{\ast}_{Kq} - A_{qK}A^{\ast}_{qq}) ~,
   \nonumber \\
   \epsilon_C &~=~& -\Im (A_{KK}A^{\ast}_{qq} - A_{Kq}A^{\ast}_{qK}) ~,
\end{eqnarray}
and $d^2 \sigma / d\Omega_\gamma d\Omega_p$ is given by
Eqs.~\ref{eq:d2sig} and \ref{eq:ABC}. Note that with three values
measured for the normal-component spin asymmetry (corresponding to the
three angles of the photon detectors), we have three expressions from
which the coefficients $\epsilon_A$, $\epsilon_B$, and $\epsilon_C$
can be determined. Although this inversion is possible in principle,
it is more efficient to use the measured asymmetries directly in our
fitting procedure to determine the independent $A_{i\mu}$'s, which we
describe in the next section.

With a normally-polarized beam, the photon detector directly above the
target could measure another piece of the normal-component coincident
analyzing power, given by
\begin{equation}
   {{d^2 \sigma} \over {d\Omega_\gamma d\Omega_p}}
   ({\bf \hat n}) A_y ({\bf \hat n}) ~=~
   2\Im (A_{KK}A^{\ast}_{Kq} + A_{qK}A^{\ast}_{qq}) ~,
\label{eq:dAyn}
\end{equation}
where
\begin{equation}
   {{d^2 \sigma} \over {d\Omega_\gamma d\Omega_p}}
   ({\bf \hat n}) ~=~ \vert A_{KK} \vert^2 + \vert A_{Kq} \vert^2
   + \vert A_{qK} \vert^2 + \vert A_{qq} \vert^2 ~.
\label{eq:d2sign}
\end{equation}

It can be shown algebraically that the ratio of Eqs.~\ref{eq:dAyn} and
\ref{eq:d2sign} can be expressed in terms of the normal-component {\em
singles\/} observables in the form
\begin{equation}
   A_y ({\bf \hat n}) ~=~ -{{(P-A_y )} \over {(1-D_{N'N})}} ~.
\label{eq:Ayn}
\end{equation}
In Ref.~\cite{SPW95} this equation was used to demonstrate that the
independently measured $(\vec p,p' \gamma)$ coincident and $(\vec
p,\vec p\,')$ singles observables were internally consistent.

Finally, the four remaining $(\vec p,p' \gamma)$ observables were
extracted from the spin asymmetries measured when the incident beam
polarization had been oriented to lie in the scattering plane.  These
are components of the longitudinal and sideways analyzing powers,
$D_{0L}({\bf \hat k})$ and $D_{0S}({\bf \hat k})$, respectively. These
quantities, which vanish identically for singles measurements, are
related to the corresponding center-of-mass asymmetries $D_{0K}({\bf
\hat k})$ and $D_{0q}({\bf \hat k})$ via a rotation in the reaction
plane through the angle $\theta_{pK}$:
\begin{equation}
   {{D_{0L}({\bf \hat k})} \choose {D_{0S}({\bf \hat k})}} ~=~
   {{\cos \theta_{pK} ~~-\sin \theta_{pK}} \choose 
    {\sin \theta_{pK} ~~~~~\cos \theta_{pK}}}
   {{D_{0K}({\bf \hat k})} \choose {D_{0q}({\bf \hat k})}} ~,
\label{eq:D0L}
\end{equation}
with $D_{0K}({\bf \hat k})$ and $D_{0q}({\bf \hat k})$ defined by
\begin{eqnarray}
   {{8\pi}\over{3R}} {{d^2 \sigma} \over {d\Omega_p d\Omega_\gamma}}
   ({\bf \hat k}) D_{0K}({\bf \hat k}) 
   &~=~& 2[\Re (A_{n0}A^{\ast}_{KK}) - \Im (A_{nn}A^{\ast}_{Kq})]
   t_{nK} ({\bf \hat k}) \nonumber \\
   &~+~& 2[\Re (A_{n0}A^{\ast}_{qK}) - \Im (A_{nn}A^{\ast}_{qq})]
   t_{nq} ({\bf \hat k}) ~, \nonumber \\
   {{8\pi}\over{3R}} {{d^2 \sigma} \over {d\Omega_p d\Omega_\gamma}}
   ({\bf \hat k}) D_{0q}({\bf \hat k}) 
   &~=~& 2[\Re (A_{n0}A^{\ast}_{Kq}) + \Im (A_{nn}A^{\ast}_{KK})]
   t_{nK} ({\bf \hat k}) \nonumber \\
   &~+~& 2[\Re (A_{n0}A^{\ast}_{qq}) + \Im (A_{nn}A^{\ast}_{qK})]
   t_{nq} ({\bf \hat k}) ~.
\label{eq:D0K}
\end{eqnarray}

The four coefficients of $t_{nK}({\bf \hat k})$ and $t_{nq}({\bf \hat
k})$ that appear in the above expressions (i.e., these four
combinations of the $A_{i\mu}$'s) can be viewed as independent
observables, and were determined in Ref.~\cite{SPW95} by a fit to the
dual sinusoidal dependence of the measured asymmetries on both the
outgoing photon direction ${\bf \hat k}$ and the orientation of the
incident proton polarization in the scattering plane. Contained in
these four observables are clear sensitivities to the relative phases
between terms in the transition amplitude corresponding to proton and
recoil nuclear polarization projections that are normal to, and
oriented in, the reaction plane; relative phases which, by virtue of
Eq.~\ref{eq:Dab}, are inaccessible via singles $(\vec p,\vec p\,')$
measurements.

\section{Minimization Procedures}

In the set of combined $(\vec p,\vec p\,')$ and $(\vec p,p' \gamma)$
observables discussed above, we have a total of 16 measured quantities
at each proton scattering angle, which can be used to determine the 11
independent quantities required to specify the complete scattering
amplitude. Although it has been shown formally \cite{Ram94} that
certain combinations of these observables can provide an analytic
solution of this problem (via matrix inversion), this method of
analysis has several disadvantages in practice. Because only
particular linear combinations of the measured quantities are used,
some statistical information is invariably lost in this method. Of
more concern is the loss of any statistical gauge of the quality or
internal consistency of the amplitude extraction procedure; because
one will always obtain an `answer,' the assignment of errors to the
$A_{i\mu}$ amplitudes becomes somewhat ambiguous.

In this work, {\em all\/} of the observables measured at each
scattering angle are used as input to a single $\chi^2$ minimization
procedure. This method not only makes use of the full statistical
information contained in the data set, but also minimizes any bias
that could be introduced through data manipulation (e.g., forming
various linear combinations of the data) prior to the actual fitting.
Thus, our approach can be viewed conceptually as the following: we
seek values for the six complex amplitudes $A_{i\mu}$ contained in
Eq.~\ref{eq:Tpp} that, when used to form the 16 combinations provided
in Section II, minimize the differences between these combinations and
the measured values of the corresponding observables. Details of the
minimization procedure we followed will be provided in the next few
paragraphs, but we first point out a few subtle features unique to
this problem. Because the amplitudes are complex, one can seek values
for either their magnitudes and phases, or for their real and
imaginary components. In this work, we carried out independent fits to
both parameter sets, and obtained (as one would hope) equivalent
results. However, it became clear that when attempting to resolve
various discrete ambiguities in the fitting results, or invoking
arguments of `smoothness' (in momentum transfer), use of magnitudes
and phases as the fitting parameters was favored. It was also
necessary to hold one phase fixed during the fitting, as the
observables are insensitive to any uniform shift in all the phases.
Because $\vert A_{nn} \vert$ was consistently one of the largest
magnitudes over the entire angular range studied, we chose to define
$A_{nn}$ to be real and positive, and thus determined (in effect) the
phase of every other $A_{i\mu}$ relative to that of $A_{nn}$.

A final choice which required careful thought was the selection of
appropriate starting values for the parameters to be fit. To avoid any
bias, we noted that the form of the singles cross section, $d\sigma /
d\Omega_p$ (Eq.~\ref{eq:norm}), reveals that the maximum allowed value
for the magnitude of any of the $A_{i\mu}$'s is constrained by
\begin{equation}
   \vert A_{i\mu} \vert \le \sqrt{d\sigma \over d\Omega_p} ~.
\end{equation}
Thus, the allowed parameter space for the magnitudes must be
restricted to this range. In our fitting, we therefore assigned
starting values for each $\vert A_{i\mu} \vert$ by using a linear
random number generator to select a number between zero and this
maximum. Similarly, starting values for each phase were chosen at
random over the allowed range
\begin{eqnarray}
   0 \le \phi_{i\mu} \le 2\pi ~.
\end{eqnarray}
By holding $\phi_{nn}$ at zero, then searching the 11-dimensional
parameter space for minima in $\chi^2$, we were able to determine the
values for the $A_{i\mu}$ that were most consistent with our entire
16-observable data set.

We now describe the actual fitting procedure in greater detail. We
begin by defining
\begin{equation}
   \chi ^2 ~\equiv~ \sum ^{16}_{j,k=1} [F_j - f_j (A_{i\mu})] \,
   W_{jk} \, [F_k - f_k (A_{i\mu})] ~,
\label{eq:chi2}
\end{equation}
where $F_j$ are the measured values for the 16 observables, and $f_j$
are the expressions given in the previous section for these same
observables in terms of the $A_{i\mu}$'s. In this equation, $W_{jk}$
is the weight matrix, given by
\begin{equation}
   W_{jk} ~=~ [\epsilon ^{-1}]_{jk} ~.
\end{equation}
Here $\epsilon$ is the full error matrix associated with the set of
observables. In the absence of any correlations among the observables,
i.e., if each of the observables had been determined independently,
$W_{jk}$ would be diagonal, with each element equal to the inverse of
the square of the error assigned to each observable,
\begin{equation}
   W_{jj} ~=~ {1 \over {\delta F_{j}^2}} ~.
\end{equation}
In this work, however, $W_{jk}$ has been generalized to include known
correlations (off-diagonal elements) between specific observables,
given their method of determination \cite{SPW95}. For example, the
values obtained for the in-plane coincident cross-section
coefficients, $B/A$ and $C/A$, were deduced from the same data set by
fitting sinusoidal functions to the measured photon yields. Thus, the
resulting coefficients of the fit are highly correlated.

The minimization procedure we employed uses a combination of
algorithms \cite{Bev69} to locate the minima of an arbitrary,
nonlinear function in an arbitrarily large parameter space.  The
$\chi^2$ function (Eq.~\ref{eq:chi2}) can be linearized around a
minimum value $\chi^2_0$ via
\begin{equation}
   \chi^2 ~=~ \chi _{0}^2 ~+~ \sum_{\mu =1}^{11} 
   \left. {\partial \chi^2 \over \partial A_\mu} \right\vert_{A_0} 
   \delta A_\mu ~.
\label{eq:linchi}
\end{equation}
In this and following equations, we have simplified our notation by
denoting all $A_{i\mu}$ with a single index, $A_\mu$, representing
either a magnitude or phase. In the above equation, $A_0$ is the {\em
set\/} of $A_\mu$ values which minimize $\chi^2$, i.e., their values
at $\chi^2 = \chi^2_0$. Thus, our minimization condition can be
defined by
\begin{equation}
   {\partial \chi^2 \over \partial A_\nu} ~=~
   \left. {\partial \chi^2 \over \partial A_\nu} \right\vert_{A_0} ~+~
   \sum_{\mu =1}^{11} \left. {\partial^2 \chi^2 \over \partial A_\nu
   \partial A_\mu} \right\vert_{A_0} \delta A_\mu ~=~ 0 ~.
\label{eq:mincond}
\end{equation}
Once initial values for the $A_\mu$'s are chosen, the parameter search
begins by solving for optimal changes in the parameters, $\delta
A_\mu$, then generating new values for these parameters via
\begin{equation}
   A_\mu ~=~ A^{\rm init}_\mu ~+~ \delta A_\mu ~,
\end{equation}
from which point the parameter search can begin again. This procedure
is repeated until a minimum in $\chi^2$ is found.

The minimization condition can be written as a matrix equation
\begin{equation}
   \alpha \, \overrightarrow{\delta A} ~=~ \overrightarrow{\beta}
\end{equation}
or
\begin{equation}
   \overrightarrow{\delta A} ~=~ \alpha^{-1} \overrightarrow{\beta}
\end{equation}
where
\begin{equation}
   \alpha_{\mu \nu} ~\equiv~ {1 \over 2} \left. {\partial^2 \chi^2 
   \over \partial A_\mu \partial A_\nu} \right\vert_{A_0}
\end{equation}
and
\begin{equation}
   \beta_\mu ~=~ -{1 \over 2}
   \left. {\partial \chi^2 \over \partial A_\mu} \right\vert_{A_0} ~.
\end{equation}

The algorithm \cite{Bev69} used to find the $\chi^2$ minimum utilizes
a gradient search in the early stages of the fitting process, which
transforms smoothly into a linearization of the fitting function as
the fit converges. This is achieved by introducing a parameter
$\lambda$ which sets the scale for the size of steps taken along the
gradient. To do so, the diagonal elements of the curvature matrix
$\alpha$ defined above are modified according to
\begin{equation}
   \alpha_{\mu \mu} \to \alpha_{\mu \mu}^{\prime} ~=~ 
   (1 + \lambda)\, \alpha_{\mu \mu} ~.
\end{equation}
Upon inspection of this equation, we note that if $\lambda \gg 1$, then 
\begin{equation}
   \delta A_\mu ~\approx~ 
   {1 \over {\lambda \, \alpha_{\mu \mu}}}\beta_\mu 
\end{equation}
and the search is approximately a gradient search. If, on the other
hand, $\lambda \ll 1$, then 
\begin{equation}
   \alpha_{\mu \nu}^{\prime} ~\approx ~\alpha_{\mu \nu}
\end{equation}
and so the function $\chi^2$ is (approximately) linearized, as
described in Eqs.~\ref{eq:linchi} and \ref{eq:mincond}. The algorithm
is designed such that the closer $\chi ^2$ gets to a minimum
(determined by the size of the changes in the fitting parameters
required to `step across' the minimum), the smaller the value of
$\lambda$ chosen, and thus a transition from a gradient search to a
linearization of $\chi^2$ is effected.

At each scattering angle, a total of 10,000 randomly chosen sets of
the $A_\mu$'s were used as starting points to this algorithm.
Approximately 95\% of the time, the algorithm was successful and
converged on a minimum in the multi-dimensional $\chi^2$ surface. It
quickly became clear, however, that the $\chi^2$ function describing
these observables contained many local minima, and the fit would often
become `trapped' in these shallower regions, rather than converging on
the `true' minimum value, $\chi^2_{\rm best}$. (This latter quantity
was defined as simply the lowest value of $\chi^2_{\rm min}$
determined in any of the 10,000 fits.) As a result, the algorithm was
able to locate minima with $\chi^2_{min} \le 2 \, \chi^2 _{\rm best}$
only about 10\% of the time, an indication of the complexity of the
space being probed. Nevertheless, this yielded a sample of roughly
1000 sets of amplitudes at each angle which gave reasonably good
descriptions of all 16 measured observables.

The procedures used to extract final values for each spin-operator
amplitude will be presented in the next section. For now, we note that
within these 1000 or so acceptable solutions, the {\em magnitudes\/}
of the amplitudes, $\vert A_{i\mu} \vert$, were generally found to be
very stable, and largely independent of the choice of starting
parameters. The values found for the corresponding {\em phases},
$\phi_{i\mu}$, on the other hand, were highly dependent on the
starting parameters, and thus exhibited much larger fit-to-fit
variations. Based on this, a second round of fitting was undertaken,
in which the starting parameters were {\em not\/} chosen at random: in
this analysis, the magnitudes were initially set to their best-fit
values, as determined from the first round of fitting, while the
phases were each stepped through all allowed values in a
multi-dimensional grid search. Specifically, each phase was assigned a
starting value between 0 and $5\pi/3$ in steps of $\pi/3$, with all
possible combinations used as starting sets.

As a final check on the robustness of the entire fitting procedure, we
used a set of theoretical amplitudes (see Sect.~VI) to generate values
for each of the observables that had been measured in
Ref.~\cite{SPW95}, thereby producing a `data set' comparable to that
obtained in the actual measurement. Using this `data' as input to our
fitting code, we were indeed able to reproduce the input amplitudes,
i.e., the algorithm could always find the correct solution. Once this
had been established, we performed a more realistic test in which the
calculated value of each observable was randomized with a
root-mean-square deviation $\sigma$ equal to a typical experimental
error bar for that observable. Using this data set, whose statistical
precision matched that of the actual measurements, we were again able
to reproduce our input amplitudes, within acceptable errors.

\section{Results of the Fitting}

After performing 10,000 fits to the data at each scattering angle, our
next step was to try to converge on a unique set of solutions for the
$A_{i\mu}$'s. At each angle, we first discarded all fits in which the
resulting $\chi^2$ minimum, $\chi^2_{\rm min}$, was more than twice
the lowest value found, $\chi^2_{\rm best}$. For the number of degrees
of freedom in our fitting function, this provided a confidence level
of 85\% that the true solution was included among the 1000 or so fits
kept at each angle \cite{Bro85}. Among these 1000 fits, though, the
solutions tended to cluster tightly around a very small number of
regions in parameter space, yielding a set of roughly 5--10 distinct
solutions for each angle. As mentioned previously, the amplitudes
extracted from these different solutions were often quite similar in
magnitude, but showed discrete (and correlated) variations in phase.

To proceed further, i.e., to select from among these few distinct sets
of solutions, it was necessary to introduce additional assumptions
concerning the angle dependence of the amplitudes. By imposing the
constraint that the amplitudes vary slowly and smoothly as a function
of momentum transfer (as do all the observables described by these
amplitudes), we were able to eliminate most of the remaining
ambiguities in the values determined for the $A_{i\mu}$'s. We will
illustrate this procedure with examples below. Before invoking such
arguments, it is important to note that as the magnitude of a complex
quantity goes through a minimum, both the real {\em and\/} the
imaginary components must pass close to zero; hence the phase will
typically change by roughly 180$^\circ$ as one passes through this
minimum. Conversely, it is difficult to produce such a large phase
shift {\em unless\/} a magnitude becomes very small.

We now examine in detail the two amplitudes $A_{nn}$ and $A_{KK}$.
Shown in Fig.~1 are the fitted magnitudes and the relative phase of
these two amplitudes, each plotted versus the center-of-mass
scattering angle. (For plotting purposes, all phases lie in the range
$-180^\circ \le \phi_{i\mu} \le 180^\circ$.) Different plotting
symbols at each angle correspond to different sets of solutions, and
have been displaced slightly in angle. The magnitudes of both $A_{nn}$
and $A_{KK}$ are large over the range $5^\circ \le \theta_{\rm c.m.}
\le 13^\circ$ for all valid solutions, and decrease smoothly with
angle. The phase difference $(\phi _{nn}-\phi _{KK})$ is close to zero
everywhere {\em except\/} for a single solution at $\theta_{\rm c.m.}$
= 8.8$^\circ$, indicated by a daggered `X', which is near
$-180^\circ$. If one assumes that neither $A_{nn}$ nor $A_{KK}$ passes
near zero in this region, then this phase, and all other fitted values
associated with this solution, are almost certainly unphysical. While
one could attempt to make this argument more quantitative, e.g., by
fitting the magnitudes of $A_{nn}$ and $A_{KK}$ with simple functions
that did or did not pass near zero somewhere in this angle range, we
feel that even a cursory examination of Fig.~1 tends to rule out the
possibility of a zero-crossing. These arguments will become even
stronger (though obviously somewhat more model-dependent) when we
compare our results to a wide range of realistic theoretical
predictions for these quantities, all of which display very smooth
variations over this range of momentum transfer.

It is interesting to note that the three quantities presented in
Fig.~1 largely determine the observable $D_{0q}$, which in the
direction ${\bf \hat k} \cdot {\bf \hat n}$ = ${\bf \hat k} \cdot {\bf
\hat K}$ = $1/\sqrt{2}$, ${\bf \hat k} \cdot {\bf \hat q}=0$ has the
value
\begin{equation}
D_{0q} \approx \Im (A_{nn}A^{\ast}_{KK}) = 
\vert A_{nn}\vert \vert A_{KK} \vert \sin (\phi_{nn} -\phi_{KK})~.
\end{equation}
Our measured value for this observable, being close to zero, drives
the phase difference to either zero or 180$^\circ$, but supplies
little additional information to the fit. Constraints on the
magnitudes of $A_{nn}$ and $A_{KK}$, and the resolution of this
180$^\circ$ phase ambiguity, must therefore be supplied by other
observables, demonstrating again the advantages of a large, diverse
data set.

By applying similar arguments to other $A_{i\mu}$'s, we were able to
eliminate most of the remaining solution sets, and arrive at a nearly
ambiguity-free determination of the magnitude and phase of each
amplitude at all angles. For most quantities, the few distinct
solutions left were sufficiently close in value that a simple average
could be taken, with errors enlarged slightly to reflect the range of
solutions included. Exceptions to this behavior occurred only at
$\theta_{\rm c.m.} = 16.5^\circ$, where two of the phase differences
exhibited two-fold discrete ambiguities that could not be resolved; an
interpretation for this will be given in the following section. In all
cases, the final values determined for the $A_{i\mu}$ magnitudes and
phases were consistent with those found in the fits which yielded the
lowest minimum for $\chi^2$, denoted here by $\chi^2_{\rm best}$.

As the final step in our analysis, the values obtained for
$\chi^2_{\rm best}$ were normalized to the number of degrees of
freedom in the fit, to yield $\chi^2_{\nu}$, a quantity which
statistically should lie close to unity. In this work, the values for
$\chi^2_{\nu}$ at $\theta_{\rm c.m.}$ = 5.5$^\circ$, 8.8$^\circ$,
12.1$^\circ$, and 16.5$^\circ$ were 0.75, 2.14, 2.17, and 6.47,
respectively. Because the fitting function should provide an accurate
description of the data (that is, one is {\em not\/} gauging the
appropriateness of the model in this case), a minimum $\chi^2_\nu$ in
substantial excess of 1 suggests an underestimate of the input error
for at least part of the fitted data set. To compensate for this, the
error determined in the fitting process for the magnitudes and phases
of each of the $A_{i\mu}$'s was artificially increased by a factor of
$\sqrt{\chi^2_{\nu}}$. This ensures that when the extracted amplitudes
are used to determine best-fit values (with errors) for the
observables, one will reproduce the measured input data within one
standard deviation, on average. Thus, we believe that the errors
quoted here for the magnitudes and phases of the $A_{i\mu}$'s provide
a realistic estimate of the true uncertainties inherent in the fits to
these data.

\section{Comparison of Amplitudes with Calculations}

It has been shown previously \cite{SPW95} that the large number of
spin observables measured for this transition provides a severe test
for any theoretical model. None of the calculations presented in
Ref.~\cite{SPW95} could describe the momentum transfer dependence of
all the observables over the entire range covered by the data. To
determine more precisely where weaknesses lie in these models, it
is useful to bring experiment `closer' to theory by comparing not the
measured observables, as was done in Ref.~\cite{SPW95}\, but by
comparing theoretical predictions for the $A_{i\mu}$ amplitudes with
values extracted directly from the data, as described here.

In this work, we will consider five different sets of predictions for
the scattering amplitude.  To carry out these calculations, many
details must be specified, such as: the method used
to generate the distorting potential, and how well this describes the
elastic scattering data; assumptions made regarding the structure of
the excited state; the extent to which medium corrections are
incorporated into the effective interaction; the handling of exchange
contributions to the interaction; etc. Here, we will examine
predictions for the $A_{i\mu}$'s derived from the {\em same\/} five
models as those to which the {\em observables\/} were compared in
Ref.~\cite{SPW95}. Extensive discussion on the content of each
calculation is presented in that paper, and the interested reader is
encouraged to refer to Ref.~\cite{SPW95} for more detail; only fairly
brief descriptions will be provided here.
 
Of the five calculations to be shown, all use free $t$-matrices for
the effective interaction, and Cohen-Kurath matrix elements
\cite{Coh65} to specify the transition to the excited state.  Two are
relativistic calculations in which both direct and exchange
contributions are included explicitly (DREX) \cite{DREX}. The
distorting potential for the incident and outgoing projectile waves,
however, is generated either ``self-consistently'' (the same
interaction that induces the $0^+ \rightarrow 1^+$ transition is also
folded with the $^{12}$C ground state transition density to produce
the distortions), or from an optical potential, with parameters fit to
p+$^{12}$C elastic scattering cross section and analyzing power
data. In all of the figures, the predictions of these two calculations
will be shown as a solid line (self-consistent) and a long-dashed line
(optical potential). Two nonrelativistic calculations also include
both direct and exchange contributions explicitly (DW81) \cite{DW81},
and also incorporate the effects of distortion using the same two
methods, i.e., either self-consistently (dotted line) or from the same
optical potential as was used in the relativistic calculation
(dot-dashed line). Finally, we will compare our values of the
scattering amplitudes to those predicted using a relativistic
calculation in which the full interaction has been parameterized in
terms of direct scattering processes {\em only}, with distorted waves
generated self-consistently (short-dashed line). Although exchange
processes are expected to contribute significantly to this reaction,
this last calculation appeared to describe the {\em observables\/}
measured for this reaction better than any model in which exchange
contributions had been explicitly included \cite{SPW95}. This would
suggest that most current methods of accounting for exchange processes
are inadequate to describe proton-nucleus scattering at intermediate
energies, at least for unnatural parity transitions.

Before making detailed comparisons between data and theory, there is
one ambiguity which must be pointed out. Because the individual
spin-operator amplitudes have dimensions of $\sqrt{\mu {\rm b} / {\rm
sr}^2}$, cross section data \cite{Com81} were used to provide an
overall normalization factor for all spin observables considered in
the fitting procedure. This ensures, for example, that
Eq.~\ref{eq:norm} is obeyed, and the sum of the squared amplitudes
equals the measured cross section. However, as can be seen in Fig.~2,
the five calculations considered here do not reproduce this cross
section. In particular, all of the calculations underpredict $d\sigma
/ d\Omega_p$ at small angles, except for the DW81 calculation using
distortions generated in an optical model, which overpredicts the data
at small angles, then decreases much more rapidly with angle than the
data. The DW81 calculation which uses self-consistent distortions, on
the other hand, is a much more shallow function of angle than the data
suggest.

In light of these discrepancies, one could consider scaling all of the
calculated magnitudes, $\vert A_{i\mu} \vert$, by a single
multiplicative factor, in order to more closely reproduce the measured
cross section. In this way, one is effectively comparing the {\em
relative\/} sizes of the spin-operator amplitudes for each model to
the measured values. This also ensures that the dominant amplitudes
will be well reproduced. On the other hand, the differences observed
between the predicted and measured cross sections might result from
particular amplitudes (especially the larger ones) being grossly over-
or under-predicted, while the calculated values for others are
actually in close agreement with the experimentally determined values.
In this case, one is best served by {\em not\/} applying an artificial
normalization, and directly comparing data and theory for each
amplitude. Because our primary goal is to identify more narrowly the
weaknesses in individual calculations, we have adopted the latter
approach here. The potential drawback, of course, is that a
calculation that is correct in all respects other than reproducing the
measured cross section will systematically miss {\em each\/} of the
extracted amplitudes by roughly the same factor.

With the above caveat in mind, we now compare the predictions of these
five models for the magnitude and phase of each amplitude to the
values deduced from our fits. The quantity $A_{n0}$, shown in Fig.~3,
is the amplitude for polarizing the recoil $^{12}$C nucleus along the
${\bf \hat n}$ direction (perpendicular to the scattering plane) with
an unpolarized incident proton. Its magnitude (upper plot) is
predicted by each calculation to be quite weak over the entire angle
range studied in this work, a feature also seen in the data.  The
differences among the various calculations are, in absolute terms,
very small. Thus, given the level of precision with which this
quantity can be determined experimentally, not much useful information
can be obtained from $\vert A_{n0} \vert$, other than confirming the
small probability for producing this particular spin
configuration. Due to the small size of $\vert A_{n0} \vert$, the
phase difference $\phi _{n0} - \phi _{nn}$, shown in the lower half of
Fig.~3, exhibits a twofold discrete ambiguity at the largest
scattering angle. As discussed earlier, if the magnitude of a complex
amplitude passes near zero, its phase can change by nearly
180$^\circ$. Our data suggest that $A_{n0}$ passes near zero (has a
local minimum) somewhere around $\theta_{\rm c.m.}$ = 16.5$^\circ$,
but our measurements can not establish this unambiguously. We also
note that at smaller angles the phase difference is relatively flat,
in agreement with all of the calculations, although the values deduced
from the data are significantly more negative (by $\sim$ 90$^\circ$)
than any of the calculations predict.

In contrast to this weak amplitude, we next examine $A_{nn}$, shown in
Fig.~4, which is the amplitude for polarizing the recoil nucleus along
the ${\bf \hat n}$ direction when the incident proton is also
polarized along ${\bf \hat n}$. The data show that $\vert A_{nn}
\vert$ is large throughout the angular range studied, and decreases
smoothly with increasing momentum transfer. All five calculations
predict this general behavior, but differ significantly in strength,
relative to the precision with which this quantity has been
determined. It is useful to note the striking similarities between
this figure and Fig.~2, the unpolarized cross section.  (For a more
quantitative comparison, one would need to square the results shown in
Fig.~4.) Just as for $d\sigma / d\Omega_p$, the values of $\vert
A_{nn} \vert$ at small momentum transfer are underpredicted slightly
by four of the calculations, and overpredicted by the DW81 calculation
which uses optical model distortions. The momentum transfer dependence
of the data is reasonably well described by the three relativistic
calculations, but is too steep (shallow) for the DW81 calculations
that use optical model (self-consistent) distortions. The inabilities
of the models to reproduce this particular amplitude are thus directly
reflected in the discrepancies found between the predictions and
measured values for the scattering cross section.

Similar behavior is seen in Fig.~5 for the other large amplitude,
$A_{KK}$, the amplitude for polarizing the recoil nucleus along the
average momentum direction using an incident proton polarized along
the same direction. The measured magnitude $\vert A_{KK} \vert$ (upper
plot), like $\vert A_{nn} \vert$, closely follows the angular
dependence of the differential cross section, and decreases
monotonically with $\theta_{\rm c.m.}$. In this case, the three
relativistic calculations {\em quantitatively\/} describe this
behavior, and agree with the data at all values of momentum
transfer. The two nonrelativistic calculations, on the other hand, do
very poorly, either over- or underpredicting the strength of this
amplitude at small angles, and predicting angle dependences which are
either too steep or too shallow, depending on the method used to
generate distortions.  Also shown in Fig.~5 is the phase difference
$\phi_{nn} - \phi_{KK}$, which is predicted by all of the calculations
to be very close to zero over this entire angular range. The data
support this idea, albeit with a fairly large statistical uncertainty,
suggesting that the two largest amplitudes contributing to this
reaction are closely matched in phase.

The third ``diagonal'' amplitude, $A_{qq}$, has some intriguing
properties. $A_{qq}$ is the amplitude for polarizing the recoil
$^{12}$C nucleus along the {\bf q} (momentum transfer) direction, with
the proton also polarized along {\bf q}, i.e., this amplitude is
associated with the spin-operator combination $({\bf \Sigma}
\cdot${\bf q})(${\bf \sigma} \cdot${\bf q}). In a meson-exchange
formulation of $NN$ scattering \cite{Sto93}, the one-pion exchange
amplitude is associated with the spin-operator $({\bf \sigma _1}
\cdot${\bf q})(${\bf \sigma _2} \cdot${\bf q}), which is very
similar. Perhaps of more significance, in a relativistic PWIA,
$A_{qq}$ is the only amplitude that contains contributions from the
pseudovector invariant $NN$ amplitude, which carries the same quantum
numbers as the pion. Because the excitation of the 15.11~MeV state is
a $\Delta T=1$ transition, one-$\pi$ and one-$\rho$ exchange should
dominate the reaction process. The potentials for $\pi$ and $\rho$
exchange interfere constructively at very small momentum transfer, but
the one-$\pi$ exchange term changes sign relative to one-$\rho$
exchange at relatively low momentum transfer \cite{Hor94}, which can
lead to large cancellations. Thus, a zero crossing in the $A_{qq}$
amplitude may be a direct reflection of $\pi - \rho$ interference in
the reaction mechanism for this transition.

Inspection of the extracted values for the magnitude and phase of
$A_{qq}$, plotted in Fig.~6, clearly indicates that this amplitude
does indeed cross zero somewhere between $\theta_{\rm c.m.}$ =
5.5$^\circ$ and 12.1$^\circ$. The magnitude $\vert A_{qq} \vert$,
which exhibits a very different $q$ dependence than either $\vert
A_{nn} \vert$ or $\vert A_{KK} \vert$, is very well described by all
of the relativistic calculations, which each predict a zero crossing
somewhere near $\theta_{\rm c.m.} \sim 9^\circ$. The data are not
described quite as well by the nonrelativistic calculation which uses
self-consistently generated distortions, and are in strong
disagreement with the other DW81 calculation.  The large experimental
uncertainty in the phase difference at $\theta_{\rm c.m.}$ =
8.8$^\circ$ supports the idea that a zero crossing occurs near this
point. The two relativistic calculations which have distortions
generated self-consistently describe the momentum transfer dependence
of both the magnitude and phase of $A_{qq}$ quite well, and thus best
describe the physics contained in this amplitude.

The last two amplitudes, the off-diagonal terms $A_{Kq}$ and $A_{qK}$,
are also intriguing. Physically, they represent the amplitudes for
polarizing the recoil $^{12}$C nucleus along either the {\bf K} or
{\bf q} direction when the proton probe is polarized along either the
{\bf q} or {\bf K} direction, respectively. In a nonrelativistic PWIA,
these two amplitudes should be identically zero for this transition,
and only become non-zero if non-local effects, such as knock-on
exchange, are explicitly included \cite{Pik85}. In a relativistic
formulation, on the other hand, $A_{Kq}$ and $A_{qK}$ include
contributions due to linear couplings between the upper and lower
components of the bound nucleon wave function, even in PWIA. Because
the lower components are momentum-dependent, the nucleon is manifestly
off-shell, and non-local effects arise ``naturally.'' Formally,
$A_{Kq}$ and $A_{qK}$ are proportional to the tensor and axial vector
pieces of the invariant $NN$ amplitude, respectively. By carrying out
the spin algebra \cite{Pik85}, these can be written in terms of the
composite spin-convection current amplitudes $\langle {\bf \sigma}
\times {\bf J} \rangle$ and $\langle {\bf \sigma} \cdot {\bf J}
\rangle$, again respectively. Thus, these two amplitudes should be
sensitive to the off-shell behavior of the nucleons inside the
$^{12}$C nucleus.

In Fig.~7, we see that the extracted magnitude of $A_{Kq}$ is
significantly smaller than all of the predicted values, especially at
small angles.  This suggests a quenching of the tensor component of
the $NN$ amplitude within the nucleus, though one can not tell if this
is a nuclear structure or an interaction effect. It is very curious
that the only relativistic calculation that does {\em not\/} include
exchange (short-dashed line) also predicts the largest magnitude for
an amplitude that, at least nonrelativistically, is driven largely by
exchange contributions. Unfortunately, little insight is provided by
the phase difference, $\phi_{nn} - \phi_{Kq}$.  Note, however, that if
the phase found at the largest angle is simply shifted by 360$^\circ$,
the general shape of the angular dependence of the phase difference is
followed reasonably well by three of the calculations, albeit with a
$\sim 90^\circ$ offset.

In striking contrast, the other off-diagonal amplitude, $A_{qK}$, is
predicted to be somewhat smaller than $A_{Kq}$ over most of this angle
range, while the experimentally determined values are seen in Fig.~8
to be quite a bit larger. Only the DWIA calculation using optical
model distortions is consistent with the typical strength suggested by
the data. Figure~8 also shows, though, that the limited statistical
precision with which this amplitude has been determined makes it
difficult to draw any conclusions about the momentum transfer
dependence of this quantity.  This limitation is also evident in the
phase difference, in which a tight clustering about zero is broken
only at the largest angle, where a two-fold discrete ambiguity is
observed at $16.5^\circ$. As was the case for $A_{n0}$, our
statistical accuracy is such that we cannot determine experimentally
whether this amplitude is crossing zero in this angle regime or
not. Two calculations predict amplitudes that, in the complex plane,
pass near zero on one side ($\Delta \phi = +180^\circ$), two pass on
the other side ($-180^\circ$), and one phase remains constant. All of
these possibilities are consistent with one of the allowed solutions
extracted from the data.

\section{Summary and Conclusions} 

We have made a model-independent determination of the full transition
amplitude for the $^{12}$C$(p,p')^{12}$C$^{\ast}$ (15.11~MeV, 1$^+$,
$T=1$) reaction at an incident beam energy of 200~MeV at four
scattering angles. This represents the first such determination for
any hadron-induced nuclear transition other than those with $J_i = J_f
= 0$. By imposing only loose ``smoothness'' constraints on the
momentum transfer dependence of the individual spin-operator
amplitudes, we have performed a nearly ambiguity-free extraction of
these quantities, which has provided deeper insight into the physics
driving this transition.

As expected theoretically, the two diagonal spin-operator amplitudes
$A_{nn}$ and $A_{KK}$ were found to be the dominant amplitudes over
the entire momentum transfer range studied. Each of these was better
described by the three calculations carried out in relativistic
frameworks than by the two nonrelativistic calculations we
considered. The third diagonal amplitude, $A_{qq}$, which has a
spin-operator structure similar to that of one-pion exchange,
exhibited behavior characteristic of a ``zero crossing.'' The value of
momentum transfer at which this occurred was again much better matched
by the relativistic calculations than the DWIA predictions.  This
crossing has physical significance in that it may reflect interference
between $\pi$ and $\rho$ exchange, and hence may serve as a gauge of
the relative strengths of these two contributions within the nuclear
medium.

The three off-diagonal amplitudes are all much weaker than the three
just discussed, and as such were determined with much larger
experimental uncertainties. The magnitude of the $A_{n0}$ amplitude
provided little insight, though the phase was consistently (i.e., at
all angles) off by about $90^\circ$ relative to most of the predicted
values. The amplitudes $A_{Kq}$ and $A_{qK}$ are expected to be
sensitive to the off-shell behavior of the nucleons inside the
$^{12}$C nucleus, and should therefore probe the non-local or exchange
nature of the scattering process. Despite the sizable errors on the
experimentally determined values for these two amplitudes, neither are
described well by any of the five calculations considered here, which
may indicate problems in our present treatment of non-local effects in
both relativistic and non-relativistic frameworks.

\bigskip
\centerline{\bf ACKNOWLEDGEMENTS}
\bigskip

We thank those who participated in the experiment from which the
observables discussed here were obtained. We also thank J.~Piekarewicz
and J.~R.~Shepard for providing us with the calculations and for
useful discussions.

This work was supported in part by the US National Science
Foundation under Grant No.\ PHY-9602872.

\newpage
\narrowtext

\begin{figure}
\caption {Sets of fitted values for the magnitudes of $A_{nn}$ {\it
(top)} and $A_{KK}$ {\it (middle)}, and their relative phase
difference {\it (bottom)}, each plotted vs.\ $\theta_{\rm c.m.}$. In
each case, different plotting symbols at each angle represent the
results of different solutions. The symbols have been displaced
slightly in angle for (some) clarity.} 
\end{figure}

\begin{figure}
\caption {Differential cross section for the
$^{12}$C$(p,p')^{12}$C$^{\ast}$ (15.11~MeV) reaction at 200~MeV. The
data are from Ref.~[13]. The five curves shown are described in the
text.} 
\end{figure} 

\begin{figure}
\caption {Magnitude {\it (upper)} and relative phase {\it (lower)} of
the amplitude $A_{n0}$ plotted vs.\ the center of mass scattering
angle. The phase difference is with respect to the $A_{nn}$ amplitude.
The error bands represent best fit solutions to the data of Ref.~[12].
The five curves shown are theoretical predictions described in the
text.} 
\end{figure}

\begin{figure}
\caption {Same as Fig.~3, but for the amplitude $A_{nn}$. In our
fitting procedures, the phase of this amplitude was defined to be
zero.} 
\end{figure}

\begin{figure}
\caption {Same as Fig.~3, but for the amplitude $A_{KK}$.}
\end{figure}

\begin{figure}
\caption {Same as Fig.~3, but for the amplitude $A_{qq}$.}
\end{figure}

\begin{figure}
\caption {Same as Fig.~3, but for the amplitude $A_{Kq}$.}
\end{figure}

\begin{figure}
\caption {Same as Fig.~3, but for the amplitude $A_{qK}$.}
\end{figure}

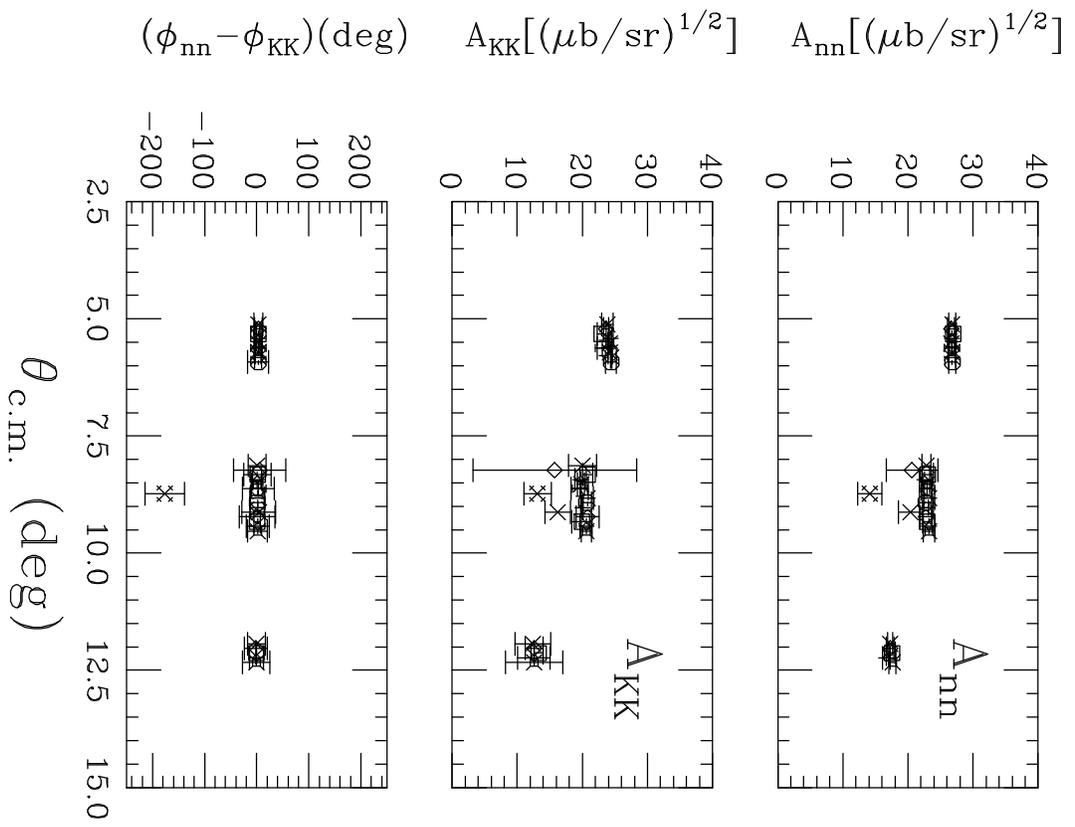
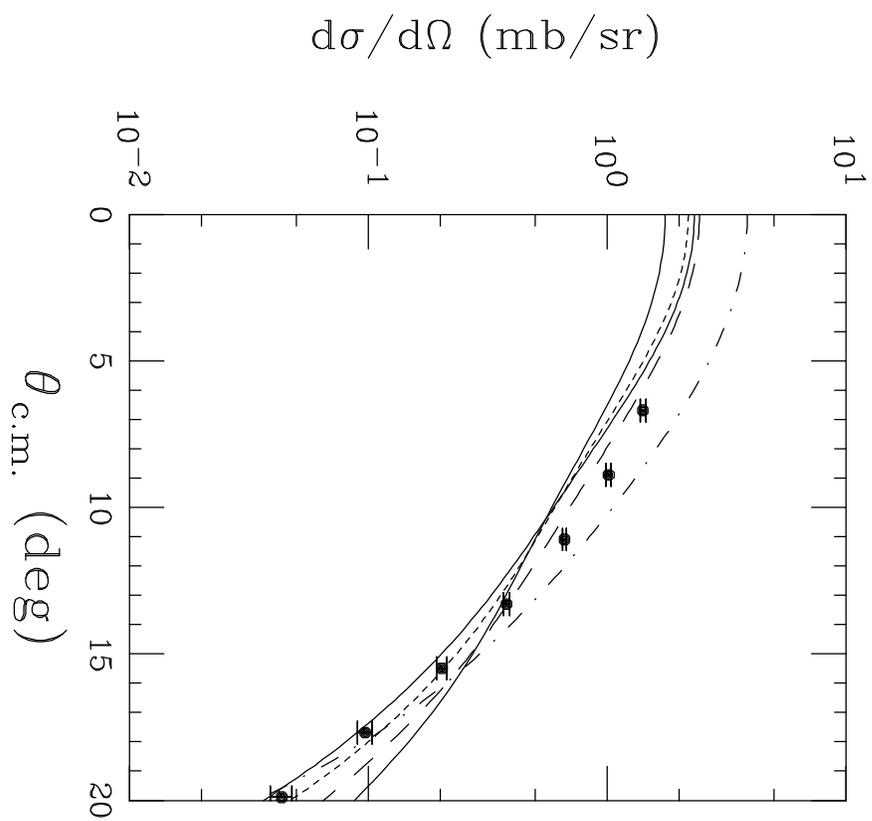
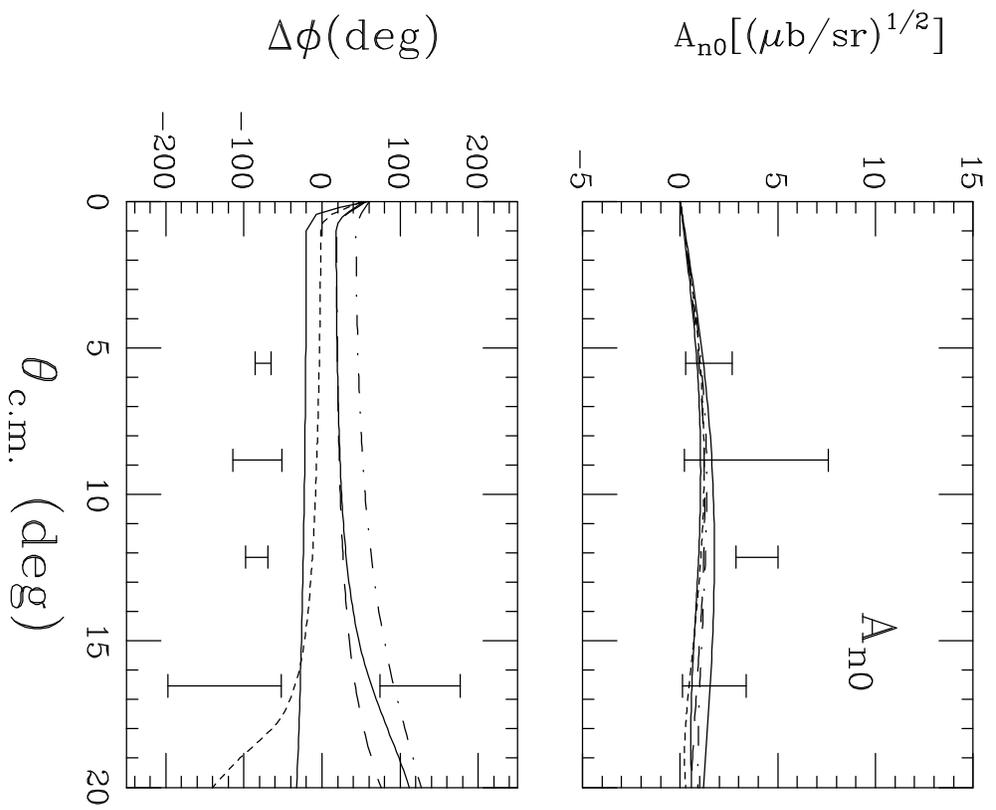
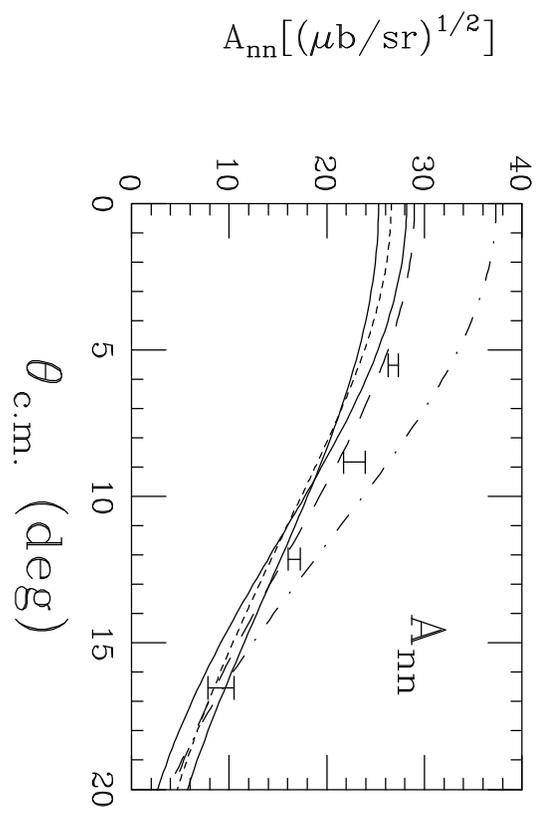
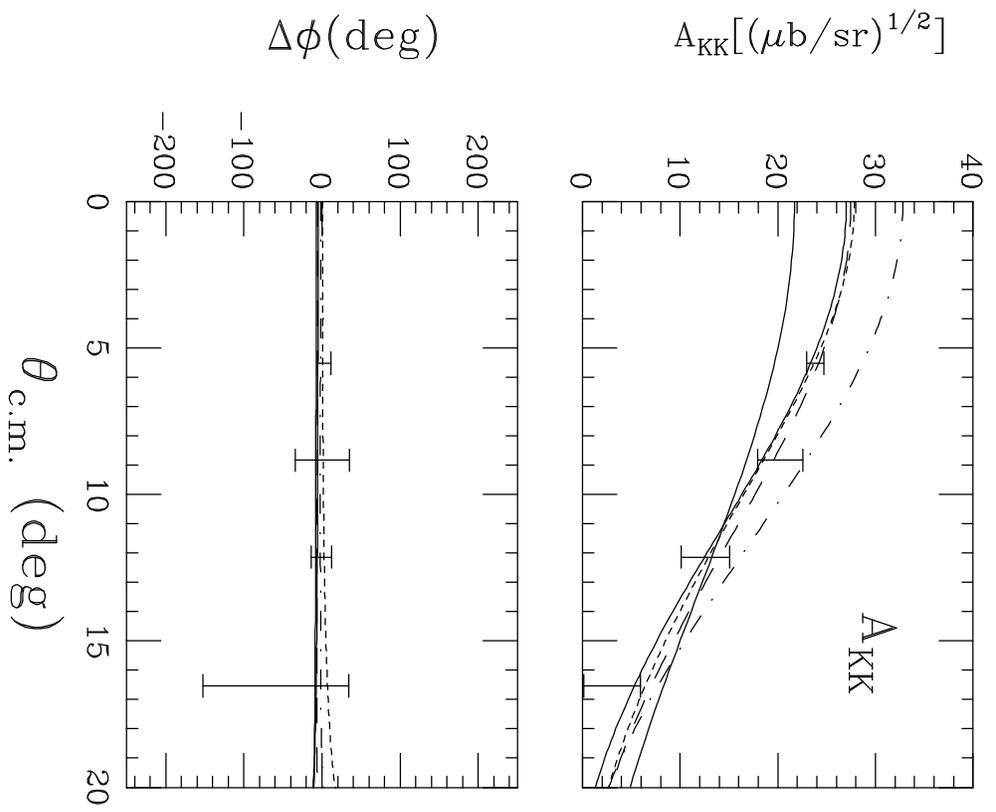
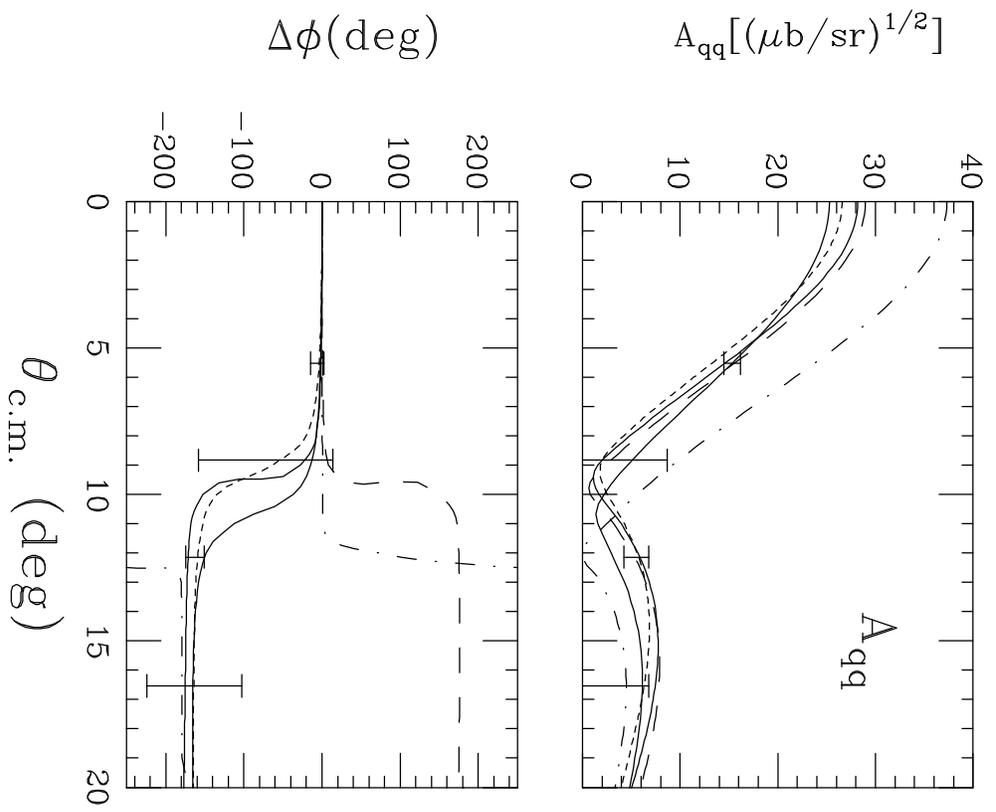
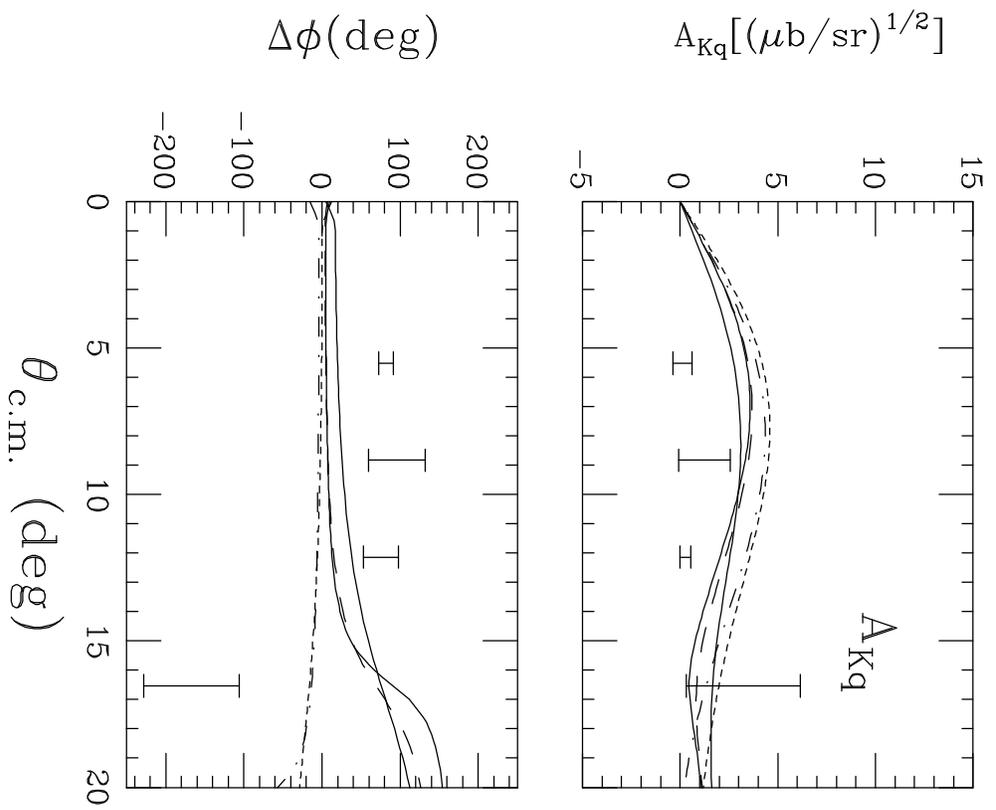
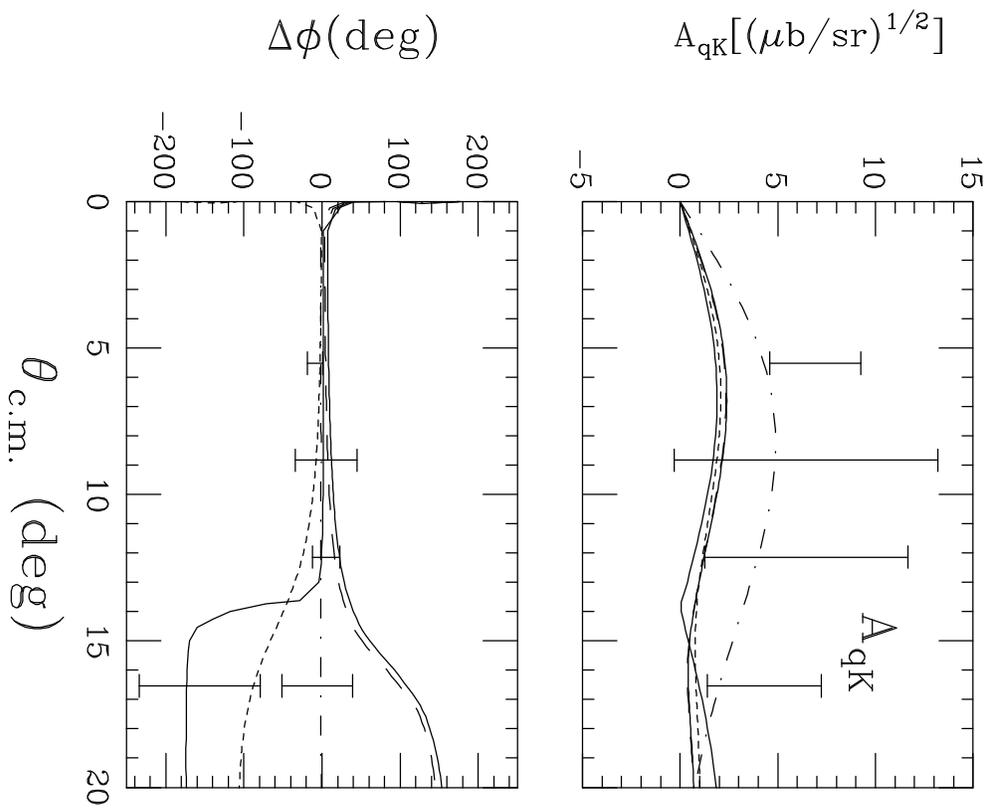
\end{document}